\documentclass[%
 reprint,
 amsmath,amssymb,
 aps,
]{revtex4-1}

\usepackage{graphicx}% Include figure files
\usepackage{dcolumn}% Align table columns on decimal point
\usepackage{bm}% bold math
\begin{document}

%\preprint{APS/123-QED}

\title{Entropy of non-local gravity}% Force line breaks 
\author{Ali Teimouri}
\email{a.teimouri@lancaster.ac.uk, ilia.teimouri@gmail.com}
 \affiliation{Consortium for Fundamental Physics,\\ Lancaster University,\\
Lancaster, LA1 4YB, UK.}

\begin{abstract}
For higher derivative theories of gravity, it is possible to write the action in terms of auxiliary fields. In such case, one can show that the equations of motion for both actions are equivalent and hence the actions themselves. In this paper we show that one can obtain the Wald's entropy from the equivalent action. We use this useful approach to localise a non-local gravitational action and calculate its associated entropy.  

\end{abstract}

\maketitle

%\tableofcontents
\section{Introduction}
Einstein's theory of general relativity can be modified in number of ways to address different aspects of cosmology. One approach to account for modified gravity in the infrared (IR) regime is the non-local gravity. Non-local gravity is constructed by inversed d'Alembertian operators that are accountable in the IR regime. Such modification to the theory of general relativity arises naturally as quantum loop effect \cite{nonloc} and used initially by \cite{Deser:2007jk} to explain the cosmic acceleration. Non-local gravity further used to explain dark energy \cite{Maggiore:2014sia}. Since such gravity is associated with large distances, it is also possible to use it as an alternative to understand the cosmological constant \cite{ArkaniHamed:2002fu}. Additionally, non-local corrections arise, in the leading order, in the context of bosonic string \cite{Aharony:2011gb}. 

It is argued in \cite{Steve} that, non-locality may have a positive rule in understanding the black hole information problem. Recently, the non-local effect was studied in the context of Schwarzschild black hole \cite{Mitsou:2015yfa,Calmet:2017qqa}. In similar manner, the entropy of some non-local models were studied in \cite{Solodukhin:2012ar}. 

In the light of these developments, in this paper, we are going to obtain the entropy for a non-local gravitational action. In this theory the non-local operators (\textit{i.e.} inversed d'Alembertian operators)
can be of any order, namely finite or infinite. The entropy is being obtained by using the Wald's description and getting use of an equivalent action, where we introduced auxiliary fields to localise the theory. 

\section{\label{sec:level1}  Higher derivative  and non-local gravity}
In this section we shall introduce two types of higher derivative actions and show  how they can be written equivalently in terms of auxiliary fields.

\subsection{\label{uveqv}Higher Derivative Gravity}
General relativity is a diffeomorphism invariant theory, thus one can add covariant higher derivatives to the action. Such modifications to the gravity can be found in 
\cite{Biswas:2005qr}. The higher derivative action can be formulated as:
\begin{eqnarray}\label{local1}
&&S_{0}+S_{1}=\frac{1}{16\pi G}\int d^{4}x\sqrt{-g}\big[R+ R F(\bar\Box)R\big],
\nonumber\\&&\text{with:} \quad F(\bar\Box)=\sum^{m}_{n=0}f_{n}\bar\Box^{n},
\end{eqnarray}
where $G$ is the Newton's gravitational constant, $R$ is scalar curvature, $\Box=\nabla_{\mu}\nabla^{\mu}$ is the d'Alembertian operator and $\bar\Box\equiv\Box/M^{2}$, this is due to the fact that $\Box$ has dimension mass squared and we wish to have dimensionless $F(\bar\Box)$, we shall note that $f_n$'s are dimensionless coefficients of the series expansion. In the above action we denoted the Einstein Hilbert (EH) term as $S_{0}$. Finally, $m$  is some finite positive integer which also can be taken to infinity as in \cite{Biswas:2005qr}. The above action can be written as \cite{Mazumdar:2017kxr},
\begin{eqnarray}
S_0+\tilde S_{1}&=&\frac{1}{16\pi G}\int d^{4}x\sqrt{-g}\Big[R\nonumber\\&+& \sum^{m}_{n=0}\Big(Rf_n\eta_n+R\chi_n(\eta_n-\bar\Box^{n}R)\Big)\Big], 
\end{eqnarray} 
where we introduced two auxiliary fields $\chi_n$ and $\eta_n$. By solving the equations of motion for $\chi_n$, we obtain: 
$\eta_n=\bar\Box^{n}R$, and hence the original action given in Eq. (\ref{local1} ) can be recovered.   
\subsection{\label{nonlocequv}Non-Local Gravity}    
It is possible to formulate a non-local action to address the IR aspects of gravity \cite{Conroy:2014eja}. The non-local action can be written as, 
\begin{eqnarray}\label{nonloc1}
&&S_0+S_{2}=\frac{1}{16\pi G}\int d^{4}x\sqrt{-g}\big[R+ RG(\bar\Box)R\big],
\nonumber\\&&\text{with:} \quad G(\bar\Box)=\sum^{m}_{n=0}c_{n}\bar\Box^{-n}.
\end{eqnarray}
\\\\
In this case the inversed d'Alembertian operators are acting on  the scalar curvature. The equivalent action can be written as, 
\begin{eqnarray}\label{nonloc2}
S_0+\tilde{S}_{2}&=&\frac{1}{16\pi G}\int d^{4}x\sqrt{-g}\Big[R\nonumber\\&+&
\sum^{m}_{n=0}\Big(Rc_n\psi_n+R\xi_n(\bar\Box^{n}\psi_{n}-R)\Big)\Big]. 
\end{eqnarray} 
Again, we introduced two auxiliary fields $\xi_n$ and $\psi_n$. Solving the equations of motion for $\xi_n$, results in having: 
\begin{equation}
\bar\Box^{n}\psi_{n}=R \quad\text{or}\quad \psi_{n}=\bar\Box^{-n}R. 
\end{equation}
Thus, the original action given in Eq. (\ref{nonloc1}) can be recovered. 

\section{\label{entropy}Entropy} 
By varying the action, it is possible to find the  N\"oether current. When the current is being conserved (\textit{i.e.} when the equations of motion are satisfied to be zero) one can define an associated potential. Now if we have a 4-dimensional space with 3-dimensional hyper surface, one can define the associated N\"oether charge by an integral over the 2-dimensional space-like boundary of the hypersurface. Wald, \cite{Wald:1993nt}, used this method and proved that the first law of the black hole thermodynamics can be satisfied when the entropy is defined in terms of a specific N\"oether charge. 
Given we have a static spherically symmetric metric, 
\begin{equation}\label{metric}
ds^{2}=-f(r)dt^{2}+f(r)^{-1}dr^{2}+r^{2}d\Omega^{2}_{2}.
\end{equation}
The Wald's entropy can be obtained as:
\begin{equation}
\mathcal{S}=-8\pi\oint \Big(\frac{\delta \mathcal{L}}{\delta
R_{rtrt}} \Big)r^{2}d\Omega^{2}_{2}=-8\pi A_H\Big(\frac{\delta \mathcal{L}}{\delta
R_{rtrt}} \Big).
\end{equation}
Where $A_H=4\pi r^{2}$ is the area of the horizon in 4-dimension.
\subsection{Case I: Higher Derivative Gravity}
The entropy for $S_0+S_{1}$ as appeared in Eq. (\ref{local1}) is given by \cite{Conroy:2015nva}: 
\begin{equation}
\mathcal{S}_0+\mathcal{S}_1= \frac{A_H}{4 G}\Big(1+2 F(\bar\Box)R\Big).
\end{equation}
Now let us calculate the entropy for $\tilde S_{1}$, 
\begin{eqnarray}
\tilde{\mathcal{S}}_1&=-&\frac{ A_H}{2 G}\times\sum^{m}_{n=0}(-\frac{1}{2}f_n\eta_n-\frac{1}{2}\chi_n\eta_n+\chi_n\bar\Box^{n}R)\nonumber\\
&=&-\frac{ A_H}{2 G}\times\sum^{m}_{n=0}(-\frac{1}{2}f_n\eta_n-\frac{1}{2}\chi_n\eta_n+\chi_n\eta_n)\nonumber\\
&=&\frac{ A_H}{4 G}\times\sum^{m}_{n=0}(2f_n\eta_n)=\frac{ A_H}{4 G}(2F(\bar\Box)R).
\end{eqnarray}
Where we fixed the lagrange multiplier as $\chi_n=-f_n$. It is clear that both $S_{1}$ and $\tilde S_{1}$ are giving the same result for the entropy as they should (see also \cite{Conroy:2015nva}). This is to verify that it is always possible to use the equivalent action and find the correct entropy. This method is very advantageous in the case of non-local gravity, where we have inversed operators.  

\subsection{\label{nonlocent1}Case II: Non-Local Gravity}
In this case calculating the Wald's entropy can be a challenging task, this is due to the fact that for an action of the form Eq. (\ref{nonloc1}), the functional differentiation contains inversed operators acting on the scalar curvature. However, by introducing the equivalent action as given in Eq. (\ref{nonloc2}), one can obtain the entropy as it had been done in the previous case. \ We know that the contribution of the EH\ term to the entropy is $\mathcal{S}_{0}=A_H/4G $. Thus we shall consider the entropy of $\tilde{S}_{2}$: 
\begin{eqnarray}
\tilde{\mathcal{S}}_2&=-&\frac{ A_H}{2 G}\times\sum^{m}_{n=0}(-\frac{1}{2}c_n\psi_n-\frac{1}{2}\xi_n\bar\Box^{n}\psi_{n}+\xi_nR)\nonumber\\
&=&-\frac{ A_H}{2 G}\times\sum^{m}_{n=0}(-\frac{1}{2}c_n\psi_n-\frac{1}{2}\xi_n\bar\Box^{n}\psi_n+\xi_n\bar\Box^{n}\psi_n)\nonumber\\
&=&\frac{ A_H}{4 G}\times\sum^{m}_{n=0}(c_n\psi_n+c_{n}\bar\Box^{n}\psi_n)\nonumber\\
&=&\frac{ A_H}{4 G}\times\sum^{m}_{n=0}(c_n(\bar\Box^{-n}R)+c_{n}\bar\Box^{n}(\bar\Box^{-n}R))\nonumber\\
&=&\frac{ A_H}{4 G}\times\sum^{m}_{n=0}(c_n\bar\Box^{-n}R+c_{n} R),
\end{eqnarray} 
where we took $\xi_n=-c_n$. Furthermore, we used the fact that $\bar\Box^{n}(\bar\Box^{-n}R)=R$, \cite{Conroy:2014eja}.  Clearly, for $n=0$, we get: \begin{equation}
\tilde{\mathcal{S}}_{2_{0}}=\frac{ A_H}{4 G}\times(2c_0R).
\end{equation}
This is an expected result, since Eq. (\ref{nonloc1}), for  $n=0,$ reduces to, 
\begin{eqnarray}
&&S_{0}+S_{2_{0}}=\frac{1}{16\pi G}\int d^{4}x\sqrt{-g}\big[R+c_{0} R^{2}\big].
\end{eqnarray}
By fixing $c_0=(6M^{2})^{-1}$, one recovers the  Starobinsky''s
model of gravity \cite{Starobinsky:1980te}. 
\subsubsection{de Sitter non-local gravity}
Non-local gravity was studied in the de Sitter (dS) universe in \cite{Elizalde:2011su}.
For dS case, the action takes the form of:
\begin{equation}
S_{dS}=\frac{1}{16\pi G}\int d^{4}x\sqrt{-g}\big[R-2\Lambda+ RG(\bar\Box)R\big],
\end{equation}
where $\Lambda=3/l^{2}$ is the cosmological constant in 4-dimension, $R=12/l^2$,  and $l$ denotes the cosmological horizon. 
Thus, the entropy would be of the form: 
 \begin{equation}
\mathcal{S}_{dS}=\frac{ A^{dS}_H}{4 G}\Big[1+\Big(2c_{0} (\frac{12}{l^{2}})\Big)\Big]=\frac{
A^{dS}_H}{4 G}\Big[1+\frac{24c_{0} }{l^{2}}\Big].
\end{equation}
The area of the horizon in dS space is given by: $A^{dS}_H=4\pi l^{2}$ in 4-dimensions.   
This  result is the same as for the \textit{infinite derivative gravity,}  \cite{Conroy:2015nva}. The matching of the results is due the fact that,  in dS background, we have constant curvatures, and thus the operators are not accountable.
\section{Summary}
We have shown that for a non-local action of the form Eq. (\ref{nonloc1}), the Wald's entropy takes the following form:
\begin{equation}
\mathcal{S}_{nonloc}=\frac{ A_H}{4 G}\Big[1+\Big(\sum^{m}_{n=0}c_{n} R+G(\bar\Box)R\Big)\Big].
\end{equation} 
It can be seen that the correction to the entropy of non-local gravity consists of two terms, one local and one non-local. We have shown  the entropy for a generic background, of the form given in Eq. (\ref{metric}),  and for the dS solution. Since, in dS background, the scalar curvature is constant; we have no contribution from non-local operators. 

Similar treatment can be done for higher order terms such as $R_{\mu\nu}G(\bar\Box)R^{\mu\nu}$ and $R_{\mu\nu\lambda\sigma}G(\bar\Box)R^{\mu\nu\lambda\sigma}$, to obtain the entropy. Yet we leave this for another study. 

The mechanism of non-locality in the context of black holes information paradox is yet not known fully.  \cite{Smolin:2011ns,Steve,Itzhaki:1995tc,Giddings:2012gc}. It would be interesting to understand the role of the entropy corrections in this context. 
\section*{Acknowledgment}
The author would like to thank Spyridon Talaganis for helpful comments.

\end{document}